\documentclass[prb,twocolumn,amsmath,amssymb]{revtex4}

\usepackage{graphicx}
\usepackage{bm}
\usepackage{multirow}
\usepackage{xcolor}
\usepackage{lipsum}
\usepackage{hyperref} 
\usepackage{newfloat}
\usepackage{siunitx}

\begin{document}

Notice: This manuscript has been authored by UT-Battelle, LLC under Contract No. DE-AC05-00OR22725 with the U.S. Department of Energy. The United States Government retains and the publisher, by accepting the article for publication, acknowledges that the United States Government retains a non-exclusive, paid-up, irrevocable, world-wide license to publish or reproduce the published form of this manuscript, or allow others to do so, for United States Government purposes. The Department of Energy will provide public access to these results of federally sponsored research in accordance with the DOE Public Access Plan (http://energy.gov/downloads/doe-public-access-plan).


\title{Complex magnetic phases in polar tetragonal intermetallic NdCoGe$_3$}

\author{Binod K. Rai$^{1,2}$} 
\email[]{binod4rai@gmail.com}
\author{Ganesh Pokharel$^{1,3}$}
\author{Hasitha Suriya Arachchige$^{1,3}$}
\author{Seung-Hwan Do$^1$}
\author{Qiang Zhang$^4$}
\author{Masaaki Matsuda$^4$}
\author{Matthias Frontzek$^4$}
\author{Gabriele Sala$^5$}
\author{V. Ovidiu Garlea$^4$}
\author{Andrew D. Christianson$^1$}
\author{Andrew F. May$^1$}
\email[]{mayaf@ornl.gov}

\affiliation{$^1$Materials Science and Technology Division, Oak Ridge National Laboratory, Oak Ridge, TN 37831, USA}
\affiliation{$^2$Currently at Savannah River National Laboratory, Aiken, South Carolina, 29808, USA}
\affiliation{$^3$Department of Physics and Astronomy, University of Tennessee, Knoxville, Tennessee 37996, USA}
\affiliation{$^4$Neutron Scattering Division, Oak Ridge National Laboratory, Oak Ridge, TN 37831, USA}
\affiliation{$^5$Spallation Neutron Source, Second Target Station, Oak Ridge National Laboratory, Oak Ridge, TN 37831, USA}

\date{\today}

\begin{abstract}
Polar materials can host a variety of topologically significant magnetic phases, which often emerge from a modulated magnetic ground state.  Relatively few noncentrosymmetric tetragonal materials have been shown to host topological spin textures and new candidate materials are necessary to expand the current theoretical models.  This manuscript reports on the anisotropic magnetism in the polar, tetragonal material NdCoGe$_3$ via thermodynamic and neutron diffraction measurements. The previously reported $H$-$T$ phase diagram is updated to include several additional phases, which exist for both $H$ = 0 and with an applied field $H \perp c$.  Neutron diffraction data reveal that the magnetic structures below $T_{N1}$ = 3.70\,K and $T_{N2}$ = 3.50\,K  are incommensurate, with a ground state  magnetic order that is incommensurate in all directions with the propagation vector $\vec{k}$ = (0.494, 0.0044, 0.385) at 1.8\,K.  A unique magnetic structure solution is not achievable, but the possible single and multi-$\vec{k}$ spin models are discussed.  These results demonstrate that NdCoGe$_3$ hosts complicated magnetic order derived from modulated magnetic moments.
\end{abstract}

\maketitle

\section{Introduction}
The search for topological states of matter is receiving significant attention in quantum materials. The realization of such phases requires fulfilling general constraints, such as specific crystal structures and magnetic ground states. Noncentrosymmetric materials are widely recognized as potential hosts for interesting and complex magnetic textures. For instance, skyrmions and solitons were realized in many $\it{chiral}$ systems\cite{Muhlbauer2009,Yu2010,Bauer2013,Qian2018, Yonemera2017, Aoki2018}. Similarly complex magnetic textures have been predicted in $\it{polar}$ systems\cite{Bogdanov2002,Keesman2016}. Polar tetragonal systems have been predicted to have cycloidal (for $C_{nv}$ symmetry) or helicoidal spin structures (for $D_{2d}$ symmetry), and these modulated ground states could give way to field-induced topological magnetic textures\cite{Bogdanov2002}.

There are many polar tetragonal materials but not all of them host field-induced topological states. Meeting the symmetry requirements alone is insufficient in this regard and further candidate materials are needed to help refine theoretical models and expand experimental efforts.  This brief summary highlights the behaviors observed in some related materials.   VOSe$_2$O$_5$ has $C_{4v}$ symmetry and exhibits a cycloidal spin state at zero-field, and a N\'{e}el-type skyrmion-lattice phase is stabilized with an applied magnetic field\cite{Kurumaji2017}. Ca$_3$Ru$_2$O$_7$ has $C_{2v}$ symmetry and was reported to exhibit metamagnetic textures\cite{Sokolov2019}. Very recently, CeAlGe with $C_{4v}$ symmetry was reported to have incommensurate multi-$\vec{k}$ magnetic phases below $T_N$, and a topological magnetic phase is claimed to be present\cite{Pupal2020}. K$_2$V$_3$O$_8$, with $C_{4v}$ symmetry, has weak ferromagnetism and a field-induced spin reorientation but the expected cycloidal spin state is not detected, though some structural complexity is also present\cite{Lumsden2001,Chakoumakos2007,Takeda2019}. Similarly, while Ba$_2$CuGe$_2$O$_7$ exhibits a helicoidal spin state consistent with its $D_{2d}$ symmetry, a topological magnetic phase has not been observed\cite{Zheludev1997,Muhlbauer2017}. It is clearly difficult to predict which materials will host topological spin textures, and thus identifying new families of candidate materials is critical in the effort to understand the relation between topologically nontrivial and trivial magnetic phases.

Rare earth based compounds display fascinating properties including a combination of topology and magnetism arising from their localized, partially filled 4f shells\cite{Nakajima2015,Hirschberger2016,Schoop2018,Hosen2018,Kaneko2019,Tabata2019,Kurumaji2019,Hirschberger2019,Seo2020}. Rare earth based intermetallic compounds of composition $RTM_3$ ($R$ = rare earth, $T$ = transition metal, $M$ = Ge, Si, Ga) crystallize in more than ten structure types\cite{Gorelenko1977,Grin1990,Moze1996,Mun2010,Goetsch2013,Subbarao2013,Arantes2018,Arantes2019}. The tetragonal crystal structure of BaNiSn$_3$ is one of the structure types that has been widely observed for $RTM_3$ materials\cite{Subbarao2013,Wolfgang1978,Parthe1983} . This crystal structure lacks inversion symmetry and is characterized by the polar space group $I4mm$ with $C_{4v}$ symmetry, and $R$ resides on a single crystallographic site.

Compounds with the BaNiSn$_3$ structure type possess a number of interesting properties such as valence fluctuations, complex magnetic ground states, heavy fermions and unconventional superconductivity\cite{Smidman2015,Kimura2005,Agterberg2007,Kawai2008,Anand2012,Adroja2012,Franz2016,Fab2016, Valenta2018}.  Here we are focusing on NdCoGe$_3$, the magnetic and transport properties of which were reported by M\'{e}asson et. al.\cite{Measson2009}. NdCoGe$_3$ was reported to undergo a paramagnetic to antiferromagnetic phase transition below a N\'{e}el temperature $T_N$ = 3.70 K, with easy-plane anisotropy revealed by measurements on single crystals \cite{Measson2009}. The material is metallic and the high-temperature susceptibility data revealed a full effective moment $\mu_{eff}$ = 3.62 $\mu_B$/Nd for Nd$^{+3}$, making this a local moment system with a Kramer's doublet ground state\cite{Measson2009}.  The nature of the spin structure was not previously investigated,\cite{Measson2009} and thus it is unknown if NdCoGe$_3$ possesses the type of modulated spin structure that might host field-induced topological phases.\cite{Bogdanov2002}

We have investigated NdCoGe$_3$ using magnetization and specific heat measurements on single crystals that were complemented by zero-field neutron diffraction experiments on polycrystalline and single crystalline samples. This phase mapping reveals a previously unreported transition at $T_{N2}$ = 3.50 K for zero-field as well as additional field-induced phases for $H \perp c$. Neutron scattering data from single crystals demonstrates that the magnetic phases below $T_{N1}$= 3.70 K and $T_{N2}$ are both incommensurate, with the ground state at 1.8 K characterized by a propagation vector $\vec{k_2}$ = (0.494, 0.0044, 0.385) that is incommensurate in all reciprocal space directions.  The possible magnetic structures obtained using this $\vec{k_2}$ and the neutron powder diffraction data at 1.8 K are discussed, including multi-$\vec{k}$ spin models. Interestingly, even multi-$\vec{k}$ models for the spin structure do not yield a constant magnitude of the moment in these modulated spin structures.  The revelation of a modulated phase at zero-field and field-induced magnetic phases suggests that NdCoGe$_3$ could be a candidate for hosting  topological spin textures as in other polar tetragonal systems\cite{Kurumaji2017,Sokolov2019,Pupal2020}.

\section{Methods}

Single crystals of NdCoGe$_3$ were synthesized from starting materials Nd (Ames Lab), Co (99.95$\%$), and Ge (99.9999$\%$) using a Ge self-flux. The starting materials were placed in Al$_2$O$_3$ Canfield crucible sets\cite{Canfield2015} in the molar ratio Nd:Co:Ge = 10:15:75 and sealed in silica under 1/3\,atm of Ar. The ampoule was heated to 1180 $^{\circ}$C and held for 6 h, cooled to 950 $^{\circ}$C over 72 h, and then the ampoule was removed from the furnace and the crystals were separated from the flux using a centrifuge.  Polycrystalline samples were made by arc-melting the elements on a water-cooled copper hearth, with the ingot being flipped and remelted several times. The as-melted ingot was then annealed for 140 h at 870 $^{\circ}$C and later ground for diffraction measurements.  The phase purity and crystal structure were examined by x-ray powder diffraction and neutron diffraction. Room temperature powder x-ray diffraction data for the ground polycrystalline material and ground single crystals were collected in a PANalytical X’Pert Pro MPD diffractometer with monochromated Cu $K_{\alpha1}$ radiation to verify phase purity and consistency.  The same diffractometer was utilized to verify the expected [001] normal orientation of the plate-shaped crystals. Specific heat $C_p (T)$ measurements were performed in a Quantum Design Physical Property Measurement System (PPMS).  The magnetization data were measured using Quantum Design Magnetic Property Measurement System and MPMS3. The ac magnetic susceptibility data were measured in the MPMS3 and PPMS.

\begin{figure}[t!]
	\includegraphics[clip,width=0.9\columnwidth]{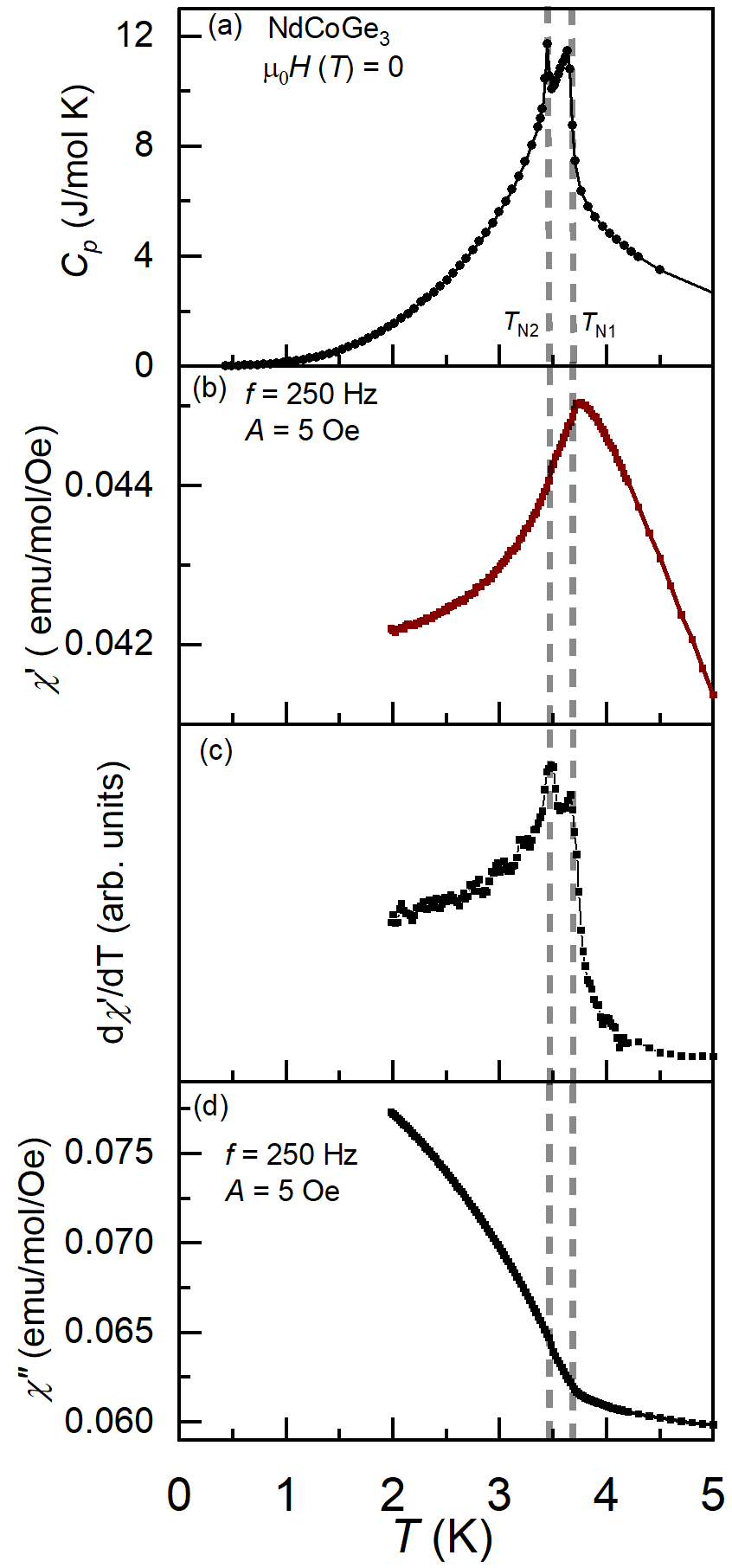}
\caption{\label{phasetransition} Zero-field data revealing magnetic phase transitions at $T_{N1}$ = 3.70 K and $T_{N2}$ = 3.50 K in NdCoGe$_3$ indicated by the vertical dashed lines.  (a) Specific heat capacity and (b,c,d) the in-plane ac magnetic susceptibility measured in zero applied DC field. The ac data are shown as the (b) in-phase contribution $\chi'$ with (c) displaying the derivative of $\chi$', and (d) contains the out-of-phase contribution $\chi''$.}
\end{figure}  

\begin{figure}[t!]
	\includegraphics[clip,width=\columnwidth]{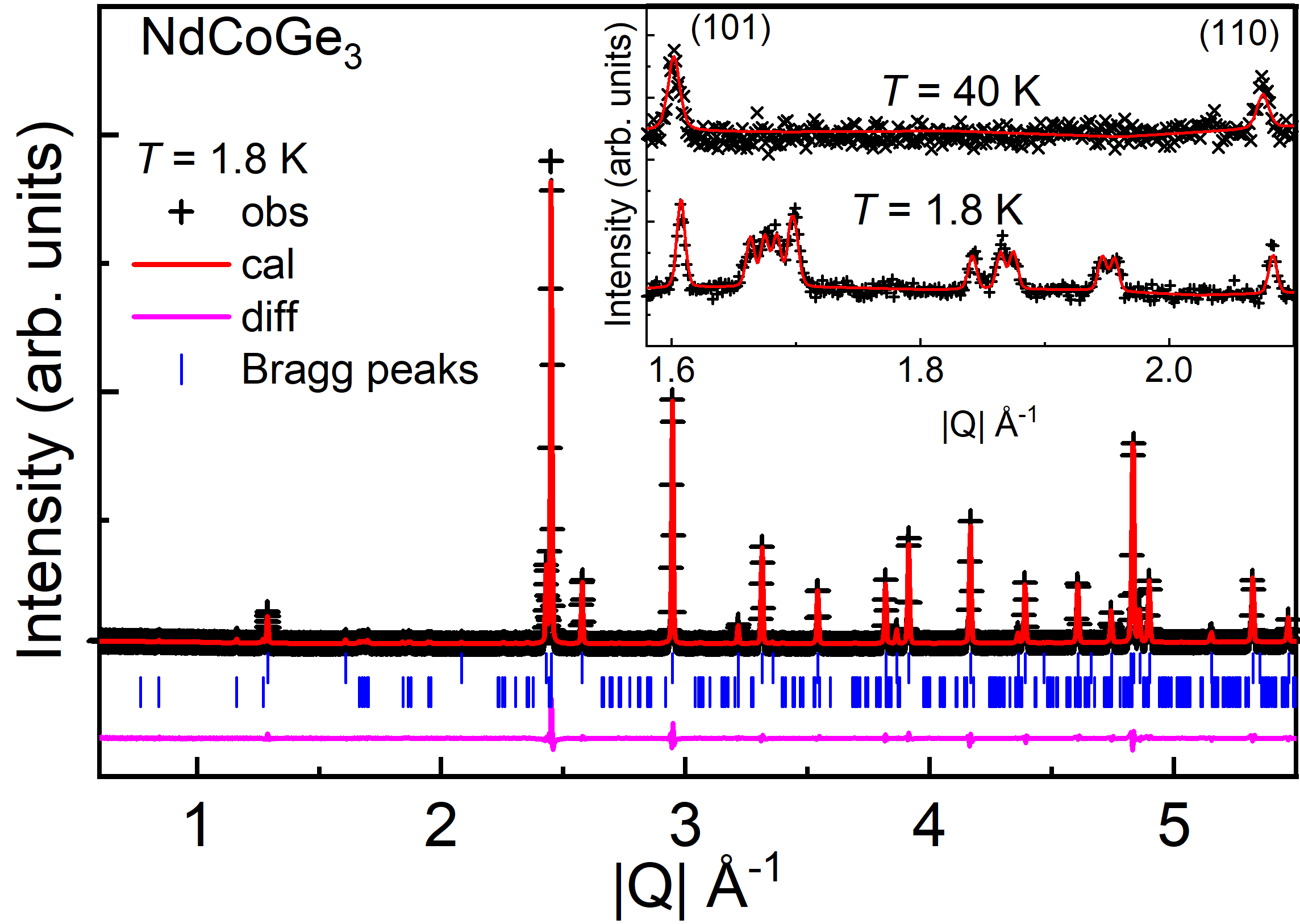}
\caption{\label{neutronzero} Neutron powder diffraction pattern of NdCoGe$_3$ at $T$ = 1.8 K $<$ $T_{N2}$ . Inset: Data for a limited $Q$ range revealing magnetic reflections at 1.8 K (bottom) through a comparison to the pattern obtained at 40 K (top) for which the Bragg peaks are entirely from the crystal lattice.}
\end{figure}  

Neutron powder diffraction data were collected on the POWGEN beamline at the Spallation Neutron Source (BL-11A, SNS) using the high resolution mode with the center of the wavelength band set to 2.665 \AA. Powders were loaded in the Al can with helium exchange gas. Neutron single crystal diffraction data were collected at PTAX (HB-1) and WAND$^2$ (HB-2C) down to $T$ = 1.5 K. A single crystal of mass equal to 104 mg was cooled down to the base temperature using a helium gas flow cryostat. Scattering planes and Bragg reflections inspected during neutron single crystal diffraction experiments were selected based on preliminary indexing of the neutron powder diffraction data.   On PTAX, PG(002) was used for the monochromator and analyzer. The horizontal collimator sequence was 48$'$-80$'$-S-80$'$-240$'$ with the neutron wavelength of 2.462 \AA. Contamination from higher-order beams was effectively eliminated using PG filters. For the PTAX experiments, the ($h$0$l$) scattering plane was used to reach the Bragg reflection 0.494,0,0.385 and scans along $h$ and $l$ were performed at various temperatures. Tilting from the ($h$, $k$, -5$h$) scattering plane allowed us to reach the Bragg reflection 0.494, 1, -2.615 and scans along $k$ were performed at various temperatures. The WAND$^2$ instrument uses the 113 reflection of its vertical focusing Ge-monochromator resulting in a wavelength of 1.486 \AA.\cite{frontzek2018} No additional collimation was used. The crystal was oriented in the $h0l$ scattering geometry and rotated in 0.1 degree steps. Each point was measured for 12 seconds. The data were reduced using the WAND$^2$ algorithms in Mantidplot \cite{Arnold2014,Mantid}.

\section{Results and Discussion}

\subsection{Magnetism in Zero Field}

The magnetic phase transitions present at zero applied field were assessed using specific heat $C_p$ and in-plane ac susceptibility measurements, the results of which are shown in Fig. \ref{phasetransition}.  The specific heat measurements revealed anomalies associated with two closely spaced phase transitions, as shown in Fig. \ref{phasetransition}(a).  The transition at $T_{N2}$ = 3.50(2) K is just below the previously reported N\'{e}el temperature of $T_{N1}$ = 3.70(2) K\cite{Measson2009}.  We observed these two anomalies in the specific heat capacity of  both single crystalline and polycrystalline samples (data not shown), which were prepared by different methods as discussed above. We do not observe any additional transitions down to $T$ = 0.4 K. The phase transitions are also observed in the ac susceptibility, which is shown as the in-phase contribution $\chi'$ and the out-of-phase contribution $\chi''$ in Fig. \ref{phasetransition}(b,d).  These transitions are most clearly evident as peaks in the temperature-derivative of $\chi'$ that is shown in Fig. \ref{phasetransition}(c).  The out-of-phase contribution shows a sharp rise on cooling below $T_{N1}$, which is not typical for a simple compensated antiferromagnet\cite{{Balanda2013}}.  The relatively large value of $\chi''$ in the paramagnetic phase is likely associated with the large electrical conductivity in this material. The ac data are for the in-plane response, with the ac drive ($A$) perpendicular to the $c$ axis.

Neutron powder and single crystal diffraction experiments were performed in zero-field to probe the nature of the magnetic order in NdCoGe$_3$.  Neutron powder diffraction data at 1.8 K are shown in Fig. \ref{neutronzero}.  The reflections originating from the crystalline lattice are indexed with the expected space group ($I4mm$) at $T$ = 40 and 1.8\,K, indicating there is no evidence of a structural phase transition upon cooling into the magnetically ordered phase(s). The refined lattice parameters at $T$ = 1.8 K are $a$ = 4.2638(1) \AA~and $c$ = 9.7507(6) \AA. 

Diffraction data for a limited range of $Q$ are shown in the inset of Fig. \ref{neutronzero} for $T$ = 40 and 1.8 K to highlight the emergence of magnetic scattering at low $T$.  The additional Bragg peaks present at 1.8 K are not allowed in the space group $I4mm$ and in fact they do not appear to be commensurate with the crystal lattice. The magnetic peaks around $\mid Q \mid$ = 1.67, 1.85, 1.94 \AA$^{-1}$ are split into multiple peaks. The peak splitting is small and approaches the instrumental resolution; for example, the separation distance between the split peaks around 1.94 \AA$^{-1}$ is only 0.014(2) \AA$^{-1}$. Due to the small splitting, unique indexing of the magnetic structure was not possible with conventional wave-vector search methods using solely the powder diffraction results.

\begin{figure}[t!]
	\includegraphics[clip,width=\columnwidth]{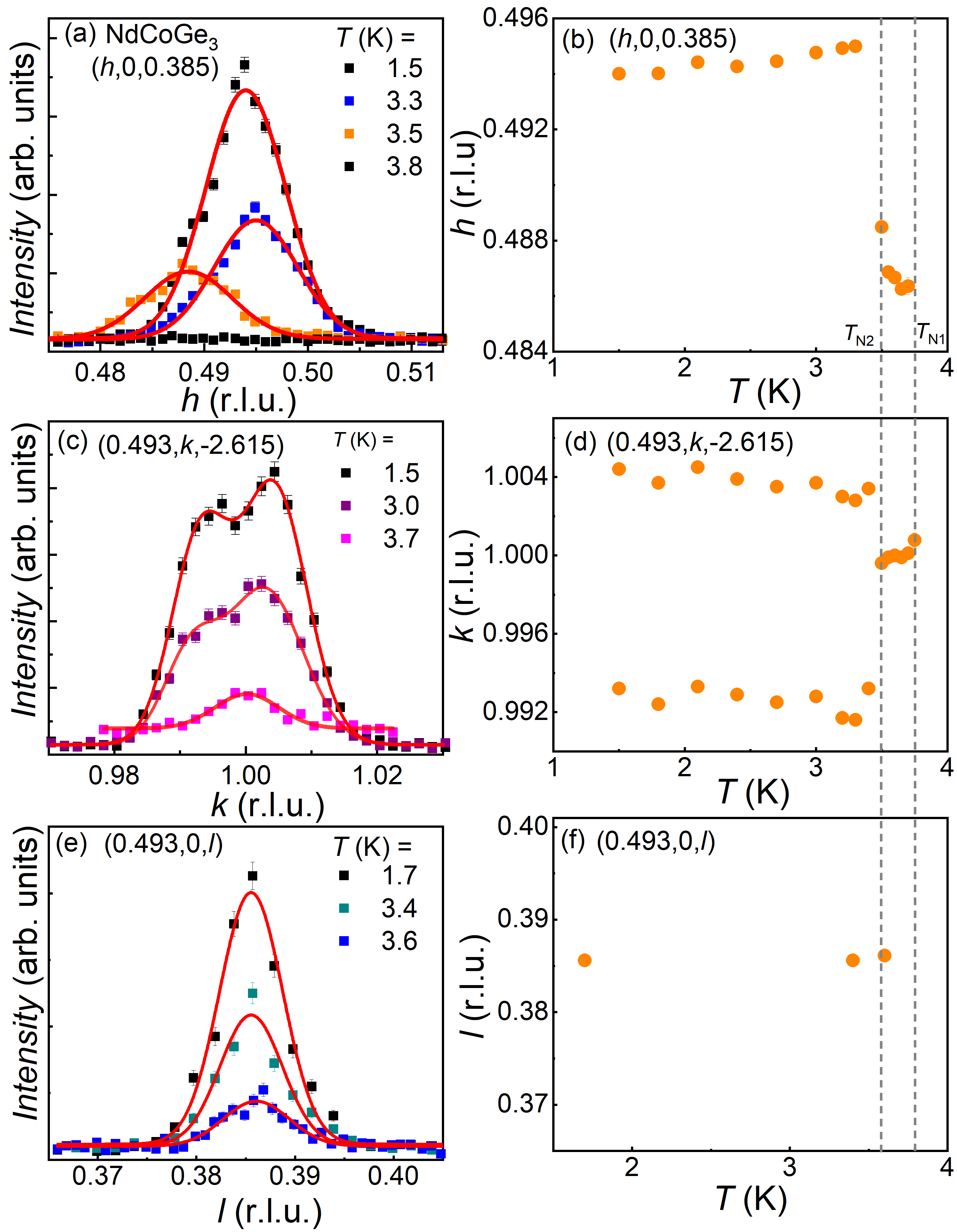}
\caption{\label{Intensity} (a,c,e) Neutron diffraction intensity as a function of $h$, $k$ or $l$ at various temperatures and (b,d,f) the fitted position of $h$, $k$, and $l$ as a function of temperature. The red lines in (a,c,e) are the corresponding Gaussian fit(s) and the fitted error bars in (b,d,f) are smaller than the size of the symbols. In (a), the small shoulders for smaller $h$ are artifacts from the sample analyzer collimator.}
\end{figure}

\begin{figure}[b!]
	\includegraphics[clip,width=\columnwidth]{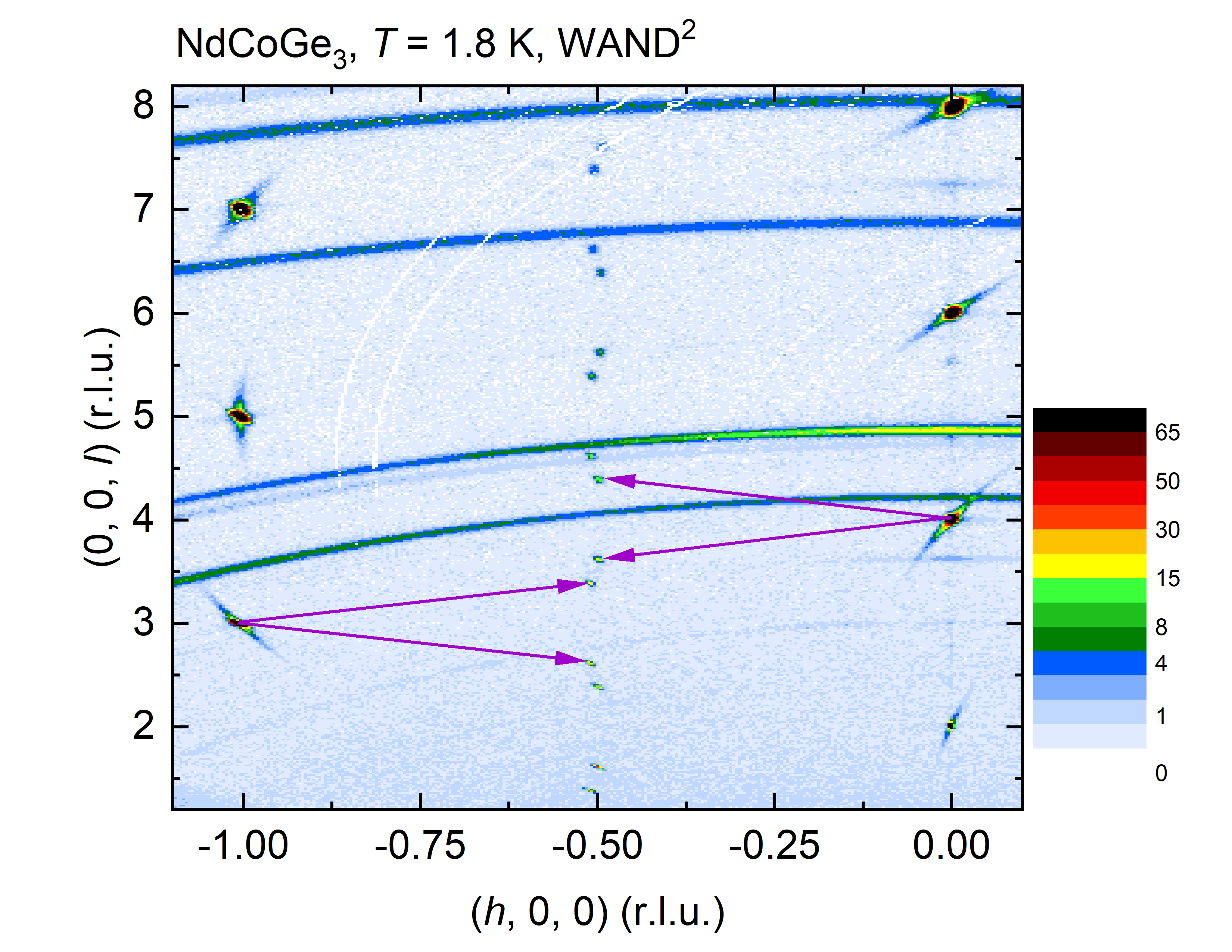}
\caption{\label{wand} Neutron single crystal diffraction data for NdCoGe$_3$ from the WAND$^2$. A section of the reciprocal $h0l$ plane between $00l$ and $-10l$ with the magnetic reflections close to the -0.5 0 $l$ position. The color scale is adjusted to show both the weak magnetic intensity and the stronger nuclear scattering. The rings originate from the Al-sample holder. Purple arrows indicate the propagation vector from the crystallographic -103 and 004 reflections.}
\end{figure}

Neutron single crystal diffraction measurements were performed to aid in determining the magnetic propagation vector(s) $\vec{k}$ and to examine the temperature dependence. The scattering planes examined were selected based on preliminary indexing of the powder data. Selected single crystal diffraction data from PTAX are shown in Fig \ref{Intensity}, and complementary data from the WAND$^2$ are shown in Fig. \ref{wand}.  In summary, the neutron single crystal diffraction data allow magnetic propagation vectors to be assigned for the different phases.   $\vec{k_1}$ = (0.486, 0, 0.385) describes the magnetic structure between  $T_{N2}$ and $T_{N1}$ (phase I in the phase diagram below) and $\vec{k_2}$ = (0.494, 0.0044, 0.385) describes the ground state at 1.8 K (phase II in the phase diagram below).  The exact value reported for $\vec{k_2}$ was obtained by refining the neutron powder diffraction data using the single crystal result as a starting point, while the value for $\vec{k_1}$ was obtained directly from the single crystal data collected on PTAX and discussed below.  Consistency between the values obtained through various approaches further confirmed the suitability of utilizing the single crystal and polycrystalline samples as complementary and equivalent, and indeed both types of data were required to convincingly obtain $\vec{k_2}$.

\begin{figure*}[t!]
	\includegraphics[clip,width=1.8\columnwidth]{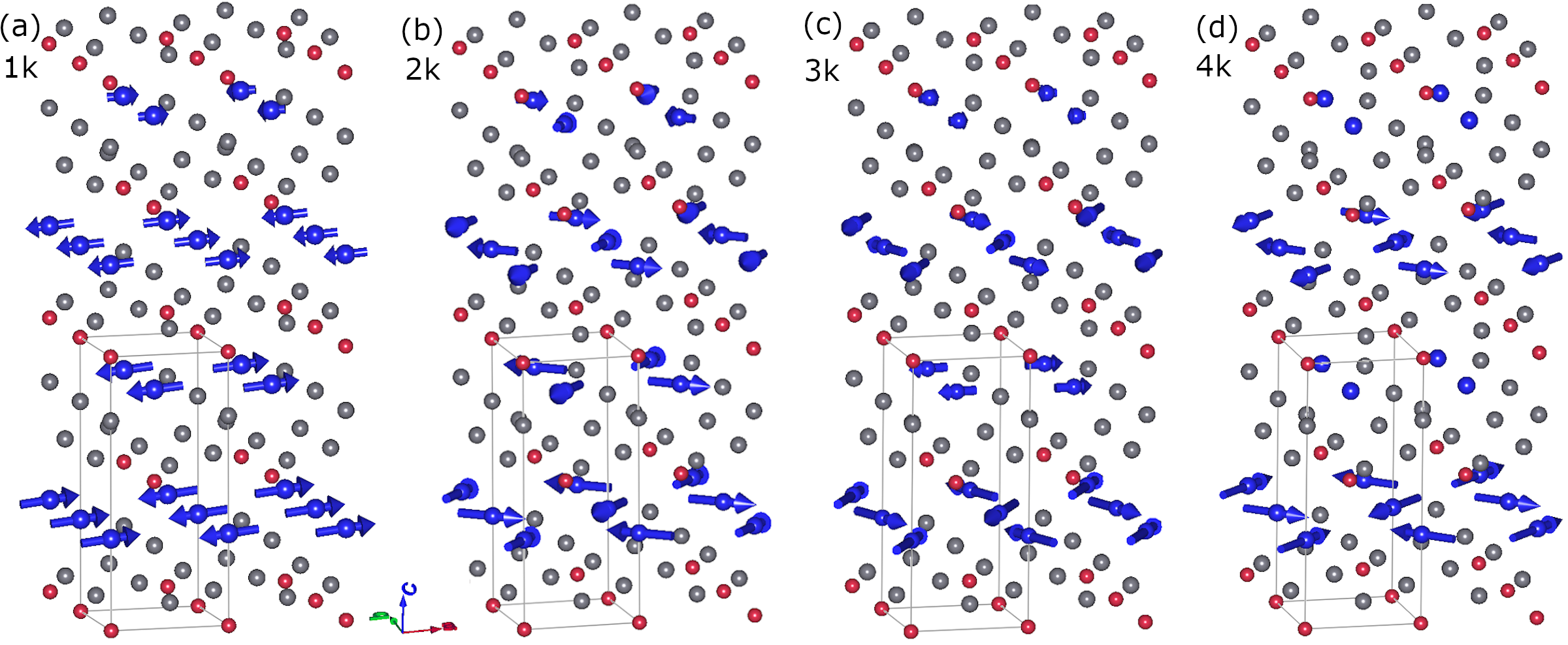}
\caption{\label{MagStructures} Candidate magnetic structures resulting from (a) 1$\vec{k}$, (b) 2$\vec{k}$, (c) 3$\vec{k}$ and (d) 4$\vec{k}$ models.  The vector moments (blue arrows) on Nd atoms (blue spheres) are shown for a segment of the magnetic unit cell; the nuclear (paramagnetic) unit cell is outlined in grey, the Co atoms are red and Ge atoms are grey.  The various models describe the neutron powder diffraction data at 1.8\,K equally well.}
\end{figure*}

Diffraction intensity while scanning along $h$, $k$ or $l$ is shown for select reflections in Fig. \ref{Intensity}(a,c,e). As shown in Fig. \ref{Intensity}(a), there is no scattering intensity at 3.8 K $>$ $T_{N1}$, consistent with a magnetic origin for this scattering.  At the base temperature of 1.5 K, all reflections are centered at incommensurate locations. The scans along $h$ and $k$ reveal fine splitting that hinders the analysis of the neutron powder diffraction data on its own.  The $h$ and $k$ components change with temperature, particularly across $T_{N2}$, whereas the $l$ component is always incommensurate with an index of $l$ = 0.385 \AA~ within error.  These trends with temperature are illustrated in Fig. \ref{Intensity}(b,d,f), where the fitted peak centers are plotted as a function of temperature.  Upon warming, an abrupt change in the $h$ component is observed near $T_{N2}$ = 3.5 K.  At the same temperature, the $k$ component collapses to the commensurate value of unity, as shown in Fig. \ref{Intensity}(d) . The asymmetrical behavior of the temperature-dependent indexing of the $k$ component about $k$ = 1 in Fig. \ref{Intensity}(d) is due to the use of a slightly tilted ($h$, $k$, -5$h$) scattering plane to reach the Bragg reflection 0.494, 1,-2.615.  The incommensurate $h$ and $l$ components in the ground state were further verified using neutron single crystal diffraction at the WAND$^2$ beamline in an $h0l$ scattering geometry at $T$ = 1.8 K. The data shown in Fig. \ref{wand} exhibit strong magnetic scattering near the 0.5 0 $l$ positions with an incommensuration in $h$ that is immediately observed in the large reciprocal space maps that can be generated using the WAND$^2$ beamline.

With $\vec{k}$ informed by the single crystal measurements, the neutron powder diffraction data were considered in detail in hopes of obtaining a spin structure for the magnetic ground state (and to refine $\vec{k_2}$ at $T$ = 1.8 K). It was determined that single and multi-$\vec{k}$ models describe the data equally well, with multiple domains required in all cases.  We now summarize the key details for the sake of discussion.  A symmetry analysis was performed using the program SARAh\cite{Wills2000} and this yielded only one possible irreducible representation for $\vec{k_2}$ = (0.494, 0.0044, 0.385) at 1.8 K. The irreducible representation consists of a linear combination of three basis vectors, permitting magnetic order along all crystal directions. From this, we analyzed the neutron powder diffraction data using the FullProf software package\cite{Rodriguez-Carvajal}. Due to the nature of the tetragonal symmetry and the propagation vector, a unique solution cannot be obtained and single $\vec{k}$ and multi-$\vec{k}$ models up to 8-$\vec{k}$ were explored.

The only single $\vec{k}$ solution is a spin density wave (SDW) that has an amplitude modulation of the magnetic moment while moving along the $c$ axis.  This model is shown in Fig. \ref{MagStructures}(a) for a partial segment of what would be a very large magnetic unit cell due to the incommensuration.  In this SDW model and in the other models discussed here, the maximum moment amplitude is 2.2(1)$\mu_B$/Nd. Our refinements did not identify any significant ordered moment at the cobalt position.  Allowing for a moment on cobalt in the SDW structure yielded moments of 0.2-0.3$\mu_B$/Co that were smaller than their error bars and the magnetic R-factor increased.  These results suggest that, if present, any moment on cobalt is below the detection limit of approximately 0.2$\mu_B$/Co (note that both Nd and Co are on equal Wyckoff positions).  In addition, we performed magnetization measurements on LaCoGe$_3$ and did not observe indications of cobalt carrying a moment in that isostructural compound.

Using the superspace formalism recently implemented for defining incommensurate modulated magnetic structures\cite{PerezMato2012}, the single-$\vec{k}$ solution can be described  by the (3 + 1)-dimension superspace group P11'($\alpha$,$\beta$,$\gamma$)0$s$.  This notation includes the standard symbol of a grey Shubnikov group, P11', followed by the propagation vector with $\alpha$, $\beta$ and $\gamma$ denoting the irrational components of $\vec{k}$, and the intrinsic translation (0$s$) associated with the point-group generator.  A 2$\vec{k}$ model can utilize $\vec{k_{2a}}$ = (0.494, 0.0044, 0.385) and $\vec{k_{2b}}$ = (0.0044, 0.494,  0.385).  This results in a non-collinear spin structure with the moments arranged in an orthogonal configuration, as shown in Fig. \ref{MagStructures}(b).  Note that an alternative  2$\vec{k}$ model can be constructed to give a spiral structure swirling around the $c$-axis. In either case the moment amplitudes remain modulated along the $c$-axis, but also inside the $ab$-plane where the amplitude change is much smoother.   In both the SDW and this 2$\vec{k}$ model, the moments are coupled antiferromagnetically along the $c$ axis.  The moments are oriented predominantly within the basal plane, though some canting along the $c$ axis may occur.  The 3$\vec{k}$ and 4$\vec{k}$ models are also non-collinear with strong moment modulations, as shown in Fig. \ref{MagStructures}(c,d).

The orientation of the moments in the candidate spin models shown in Fig. \ref{MagStructures} is consistent with the crystal electric field (CEF) model developed by M\'{e}asson et. al. on the basis of magnetization data.\cite{Measson2009}  We utilized their reported eigenvalues to compute the CEF-induced anisotropy and obtained a moment in the $ab$ plane of 2.21$\mu_B$/Nd and a moment along [001] of 1.25$\mu_B$/Nd.  These values reveal the easy-plane type anisotropy introduced by the CEF levels but show that a component along [001] may also exist.  In addition, the calculated in-plane moments are essentially equal to the maximum amplitude of the refined moments.  

The symmetries present yield an interesting case where a modulated spin structure with a constant moment (cycloidal, helical) cannot be produced using multi-$\vec{k}$ modeling up to even an 8$\vec{k}$ model (where all 8 arms of the wave-vector participate in the actual spin arrangement), for which the implementation is admittedly very difficult.  Thus, at present, it is not possible to determine the absolute magnetic structure in NdCoGe$_3$, though it is clear that the magnetic structures are complex with the ground state incommensurate in $h$, $k$ and $l$ and the moments are modulated. 

Modulated magnetic structures, with or without an amplitude of the magnetic moment, have been reported in many rare earth based intermetallic compounds\cite{Lebech1987, Cho1995, Kawano1998,Maurya2016,Zhao2017,Onimaru2019}. Due to the presence of a Kramer's doublet with local moment behavior and a lack of Kondo interaction, one would suspect that a SDW ground state is unlikely in NdCoGe$_3$.  In fact, a Kramer's doublet is expected to drive a SDW (amplitude-modulated) magnetic structure into either a commensurate magnetic structure or a modulated structure with a constant amplitude as the temperature is lowered\cite{Morin1985, Gignoux1991}.  However, a constant moment solution of the neutron diffraction data is not allowed based upon the analysis presented above. Some compounds in this family with Kramer's ions and no Kondo interactions have been reported to possess magnetic structures with a constant moment\cite{Kumar2010,Kumar2012,Fab2016,Ryan2016,Tartaglia2019}. For example, EuNiGe$_3$ has two magnetic phase transitions at zero-field and the lowest temperature phase is believed to be a modulated phase with a constant moment\cite{Fab2016,Ryan2016}. The magnetic structures of a large number of compounds in this family of materials have not yet been determined\cite{Giraldo2015,Bednarchuk2015,Nallamuthu2016,Nallamuthua2017, Klicpera2019}.

\subsection{Field-induced magnetic phases}

\begin{figure}[t!]
	\includegraphics[clip,width=0.9\columnwidth]{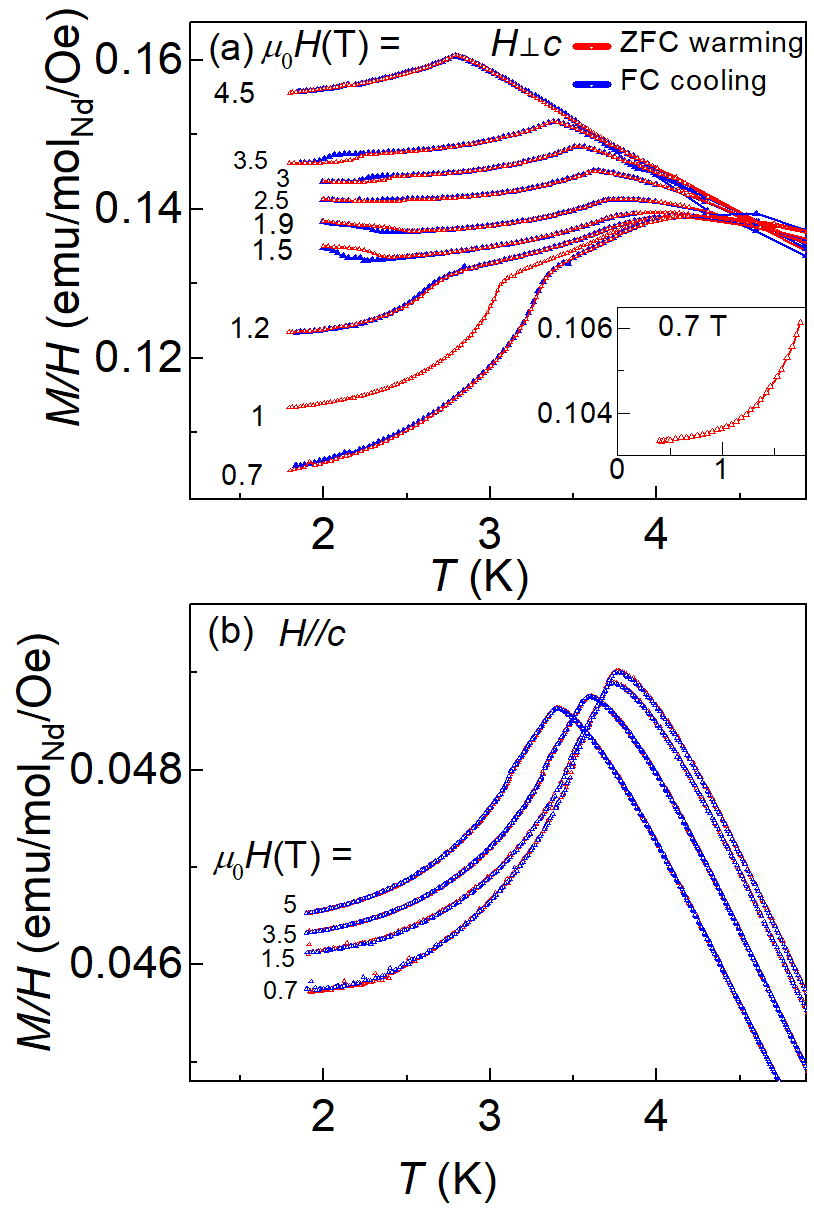}
\caption{\label{MT} Temperature-dependent magnetization in zero-field-cooled warming (red) and field-cooled cooling (blue) measurements with an applied field (a) $H \perp c$ and (b) $H||c$. The inset in (a) shows the  magnetization to $T$ = 0.4 K using the same units as the main panel.}
\end{figure}

\begin{figure}[t!]
	\includegraphics[clip,width=0.8\columnwidth]{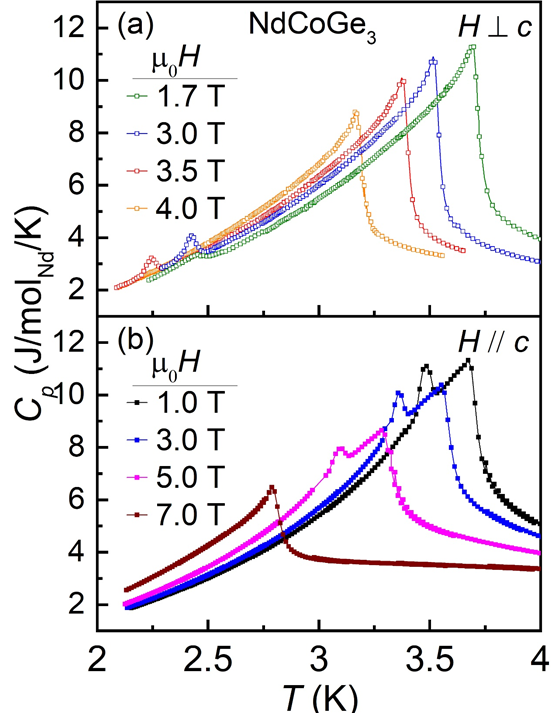}
\caption{\label{Cp} Specific heat capacity of NdCoGe$_3$ for (a) magnetic field within the basal plane and (b) field applied along the $c$ axis.}
\end{figure}

The response of the magnetism to an applied field was studied using dc magnetization $M$ and specific heat measurements. Magnetization measurements were performed down to $T$ = 0.4 K in applied fields up to $\mu_0 H$ = 6 T for both in-plane and out-of-plane orientations.  Temperature-dependent data in various applied fields are shown in Fig. \ref{MT}, where both zero-field-cooled warming (ZFC, red data) and field-cooled cooling (FC, blue data) are shown.  Examination of the data reveals a more complex response to applied field for $H\perp c$ in Fig. \ref{MT}(a) than for $H||c$ in Fig. \ref{MT}(b).  For $H||c$, the transitions at $T_{N1}$ and $T_{N2}$ are continually suppressed with applied field.  The specific heat capacity was utilized to confirm the field dependence of the magnetic phase transitions and the results are consistent with the magnetization data.  Specific heat data with a magnetic field applied within the basal plane are shown in  Fig. \ref{Cp}(a) and data for a field applied along the $c$ axis are shown in Fig. \ref{Cp}(b).

\begin{figure}[t!]
	\includegraphics[clip,width=0.9\columnwidth]{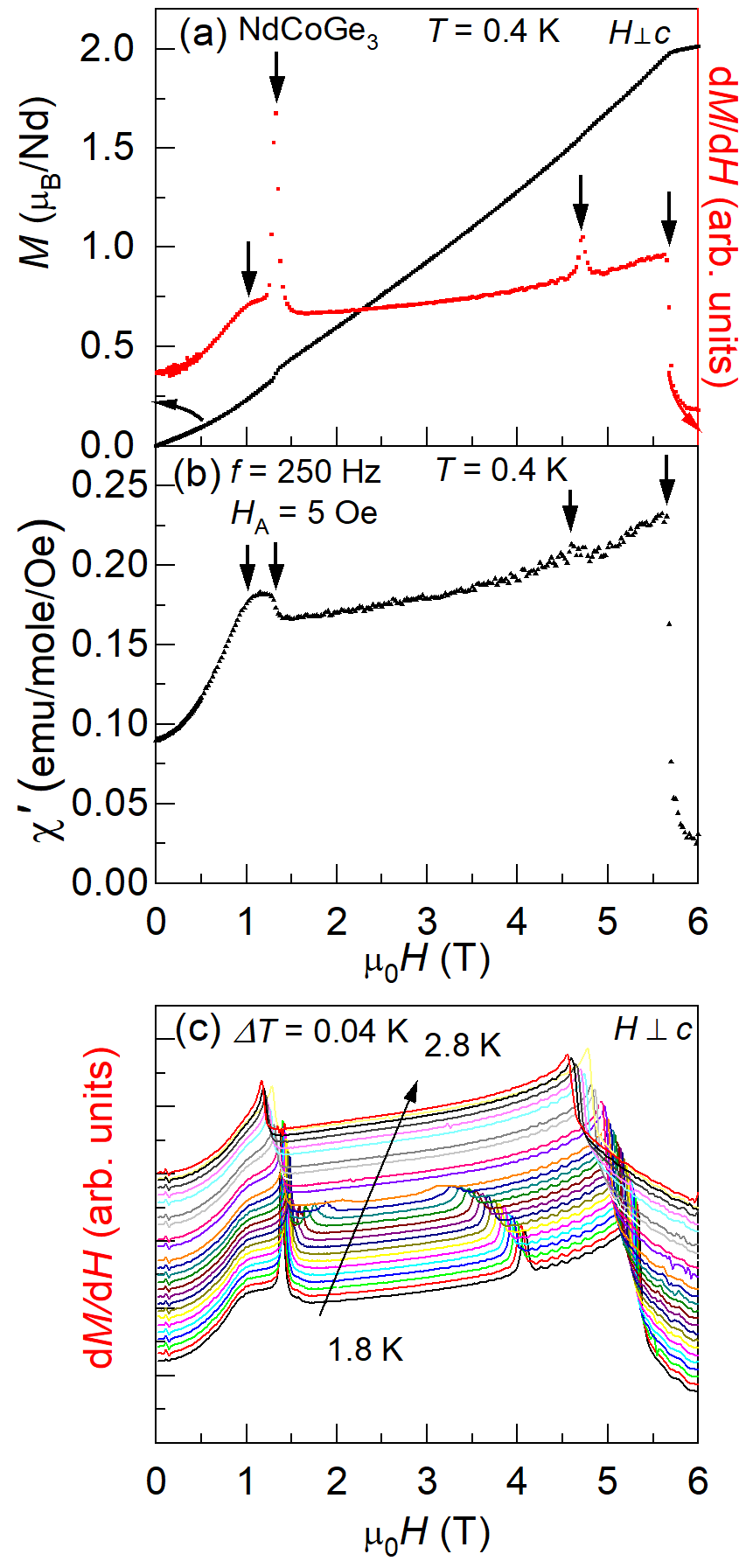}
\caption{\label{FieldM} Isothermal magnetization $M(H)$ and derivative of isothermal magnetization d$M$/d$H$ of NdCoGe$_3$ with an applied field $H\perp$c. (a) $M(H)$ (left axis) and d$M$/d$H$ (right axis) at $T$ = 0.4 K (b) The in-plane ac magnetic susceptibility $\chi$' as a function of applied DC field at $T$ = 0.4 K (c) d$M$/d$H$ from 1.8 K to 2.8 K with a temperature step of $\Delta T$ = 0.04 K. (d) d$M$/d$H$ at $T$ = 3 K.}
\end{figure} 

For $H \perp c$, the transition at $T_{N2}$ is suppressed more rapidly with increasing $H$ than is the transition at $T_{N1}$. As shown in Fig. \ref{MT}(a), the transition at $T_{N1}$ evolves from a broad feature in $M$($T$) to a sharper, cusp-like feature with increasing field.  The nature of the transition at $T_{N2}$ also evolves with increasing field, and signatures of this phase transition are not observed above 3.5 T in the data displayed in Fig. \ref{MT}(a). 

For the intermediate fields of 1.5 $\leq$ $\mu_0H$ $\leq$ 3.5 T within the basal plane, a divergence is observed between data collected in ZFC-warming and FC-cooling conditions for $H\perp c$.  No such divergence was observed for $H||c$.  The ZFC-FC divergence suggests that domains impact the magnetization.  We note that in all cases where ZFC-FC divergence is observed, the ZFC warming data (red curves) have an anomaly at a temperature larger than that observed in the FC cooling condition.  This points to thermal hysteresis associated with a first order phase transition, though the different ZFC/FC conditions clearly impact the net magnetization value as well.   In addition, the data for $\mu_0H$ = 2.5 T appears somewhat special because it contains negligible divergence.  Data for applied fields greater or less than 2.5 T have inverted magnitudes of $M$ in the region of divergence.  For instance, for $\mu_0H$ = 1.5 T the ZFC-warming data have a magnetization larger than the FC-cooling (blue curves), but the opposite is true for $\mu_0H$ = 3.5 T where the ZFC-warming data have a smaller induced magnetization than the FC-cooling data.  Considering these factors and the rising $M$ upon cooling near 2.5 K for 1.5 $\leq$ $\mu_0H$ $<$ 2.5 T, it seems that there is a non-compensated magnetic structure (small net moment) in this region of the phase diagram.  For 2.5 $<$ $\mu_0H$ $\leq$ 3.5 T, though, there is a decreasing $M$ upon cooling through the transition just below 2.5 K, which suggests that the moments are compensated (perhaps tilted and not canted). This qualitative change in the sign of d$M$/d$T$ in the region of ZFC-FC divergence could also relate to spin reorientation.

To better assess phase transitions that may be difficult to capture in $M$($T$) data, we performed isothermal magnetization $M(H)$ measurements for several temperatures with $H~\perp~c$.  We first discuss the data collected at the base temperature of $T$ = 0.4 K. Data collected upon increasing the field from a ZFC condition at $T$ = 0.4 K are shown in Fig. \ref{FieldM}(a) and the derivative d$M$/d$H$ is also shown (right axis).  Strong anomalies associated with field-induced transitions are observed in the derivative near 1.4 T, 4.7 T and 5.6 T and a weaker, shoulder-like anomaly is observed near 1.0 T (indicated by arrows).  Field-dependent ac magnetic susceptibility measurements further support the existence of these transitions as shown in Fig. \ref{FieldM}(b).  As shown in Fig. \ref{FieldM}(c), the anomaly near 1.4 T splits into two peaks that move apart in field upon warming.  The anomaly near 4.6 T at 1.8 K shifts to lower fields with increasing $T$, forming something of a dome upon merging with a split peak from lower field. The lowest and highest field anomalies are much less sensitive to the applied field than are the transitions at the intermediate fields. For the data in Fig. \ref{FieldM}(c), the crystal was cooled to 1.8 K in zero field, magnetization data were collected upon increasing and then decreasing the applied field, and then the sample was warmed to the next temperature (1.84 K) and data were again collected starting at zero field.

\begin{figure}[t!]
	\includegraphics[clip,width=0.8\columnwidth]{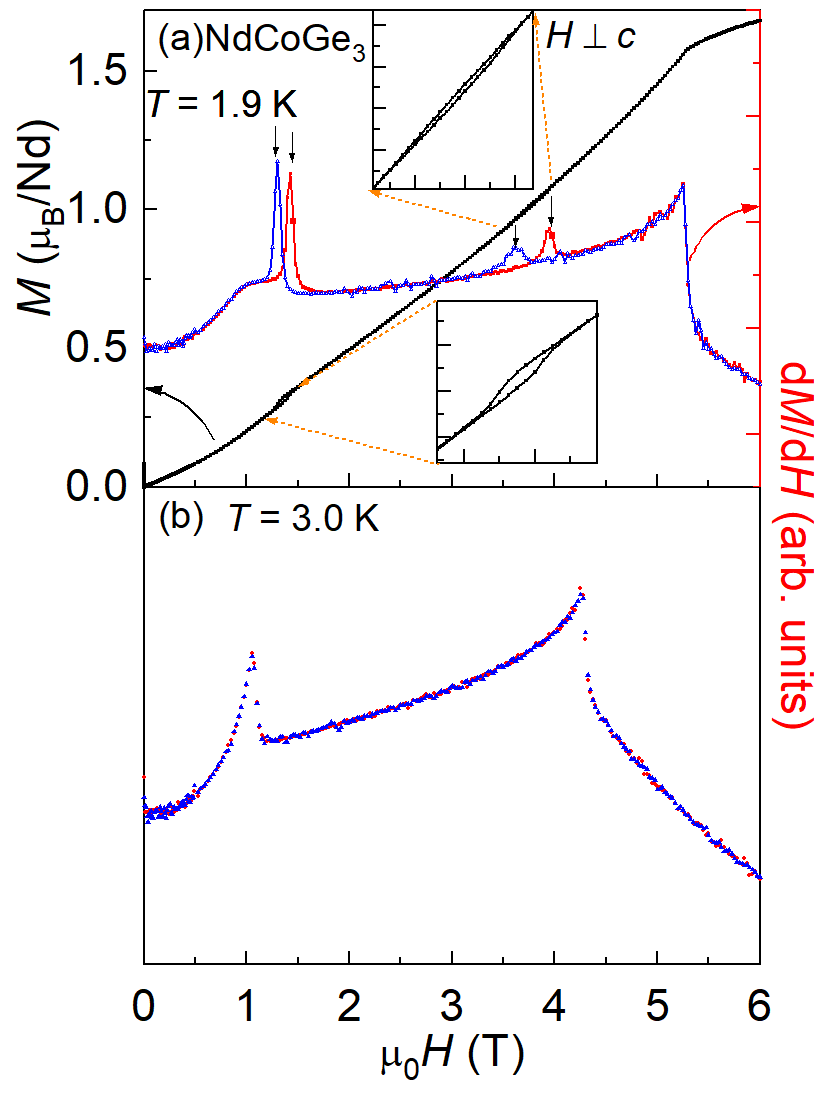}
\caption{\label{FieldH} (a,b) Isothermal magnetization data (left axis) and derivative d$M$/d$H$ (right axis) on increasing field (red line) and decreasing field (blue line) of NdCoGe$_3$ with an applied field $H\perp$c.}
\end{figure} 

The $M(H)$ data revealed several field-induced transitions for $H \perp c$, and anomalies consistent with these phase boundaries were also observed in the temperature-dependent magnetization and specific heat data discussed above.  The likely order of the various phase transitions is now discussed.   Thermal hysteresis was not observed for $H$ = 0 data (specific heat, ac susceptibility) suggesting that second order phase transitions exist at the zero-field limits of $T_{N1}$ and $T_{N2}$.  However, the discontinuity in $h$ and $k$ indexes at $T_{N2}$ as observed by neutron diffraction may imply a first order transition.  For the intermediate fields of 1.5 $\leq$ $\mu_0H$ $\leq$ 3.5 T, thermal hysteresis was observed for the ZFC-warming and FC-cooling data, suggesting a first order phase transition is present.

\begin{figure}[t!]
	\includegraphics[clip,width=0.9\columnwidth]{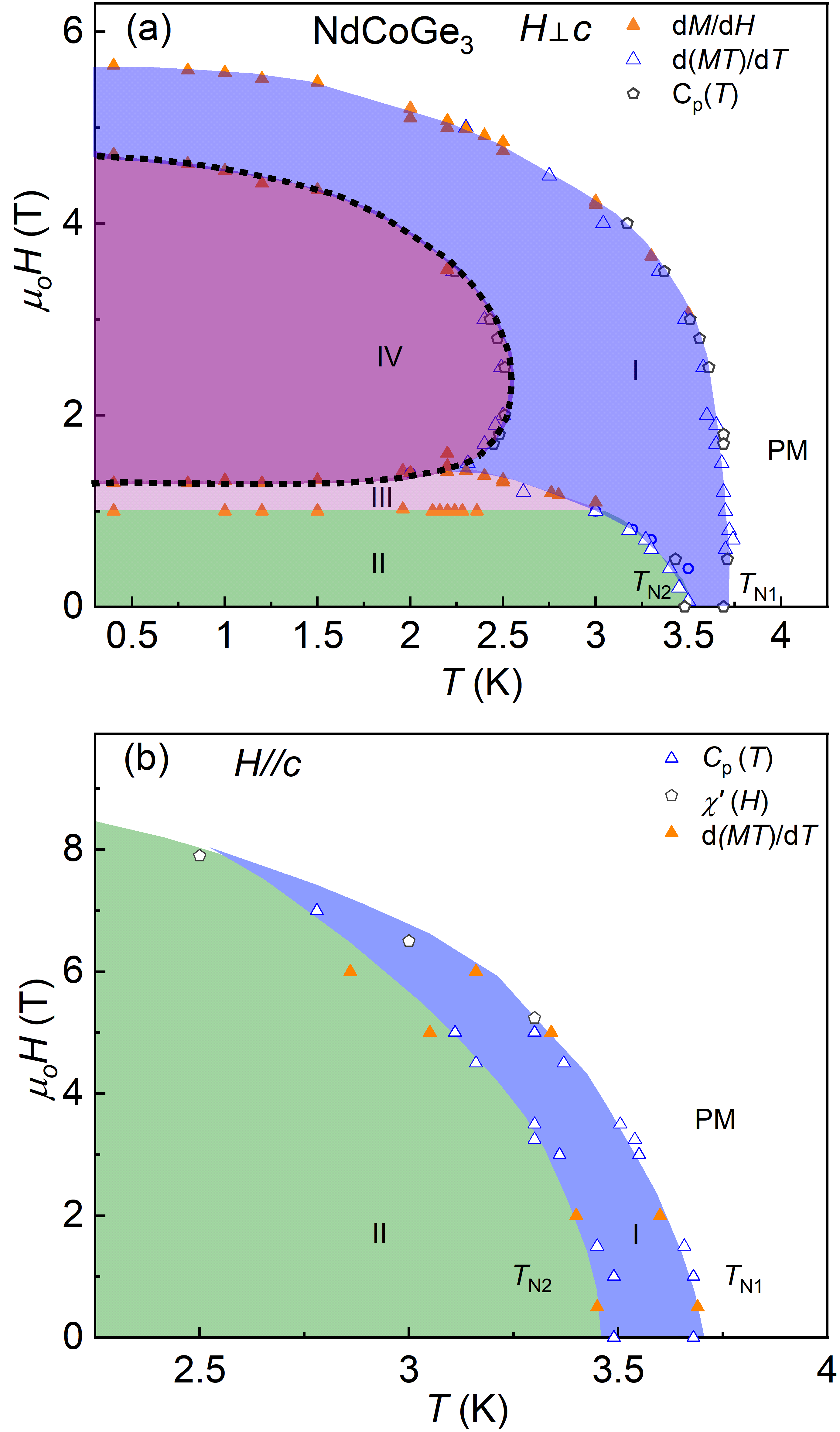}
\caption{\label{Phase} Magnetic phase diagrams of NdCoGe$_3$ as inferred from anomalies in the specific heat, derivatives of temperature-dependent and isothermal magnetization as indicated in the legends for (a) $H\perp$c  and (b) $H||$c.  The black dotted line in (a) indicates the presence of a first order phase boundary based upon observations of thermal or field hysteresis.}
\end{figure}

Field hysteresis was observed across certain phase transitions when an in-plane magnetic field was applied. Representative $M$($H$) data are shown in Fig.\ref{FieldH}(a,b) for data collected on increasing (red d$M$/d$H$ curve) and decreasing (blue d$M$/d$H$ curve) the magnetic field.   The low-temperature $M(H)$ data reveal hysteresis about the transitions at 1.4 T and 3.8 T for $T$ = 1.9 K, as shown in the insets of Fig.\ref{FieldH} (a), suggesting that these are first order transitions.  The hysteresis is clearly present in the peaks observed in the corresponding d$M$/d$H$ data. Hysteresis is not observed for the anomalies detected near 1 T and 5.2 T at 1.9 K, indicating that they are likely second-order transitions.   Above approximately $T$ = 2.5 K,  hysteresis is not observed in the $M$($H$) data, as illustrated by the data at $T$ = 3 K in Fig. \ref{FieldH}(b).  The $H$-$T$ transitions where hysteresis was observed seem to align nicely with the observation of divergence and  hysteresis in the ZFC-FC magnetization data for the same intermediate fields and low temperatures, which was illustrated in Fig. \ref{MT}(a).

The magnetic phase diagrams shown in Fig. \ref{Phase} were constructed using the magnetization and specific heat data discussed above. $T_{N1}$ and $T_{N2}$ are very close to each other, thus the lower-temperature transition was missed in zero-field measurements in the previous report\cite{Measson2009}. The magnetic phase I below $T_{N1}$ is described by the $\vec{k_1}$ = (0.486, 0, 0.385) and the phase II below $T_{N2}$ is described by $\vec{k_{2}}$ = (0.494, 0.0044, 0.385) for $H$ = 0.  The magnetic phase diagram for $H \perp c$ shown in Fig. \ref{Phase}(a) contains the field-induced magnetic phases III and IV, contrasting with the phase diagram for $H||$c shown in Fig.\ref{Phase}(b). The boundary surrounding phase IV (black dotted line) in Fig.\ref{Phase}(a) appears to involve first-order phase transitions based on hysteresis in the relevant magnetization data. Note that topologically nontrivial phases such as skyrmion lattices are often separated from trivial phases by a first-order phase boundary\cite{Neubauer2009}.  The lower critical field boundary of phase III is constructed from the shoulder feature in d$M$/d$H$. The existence of phase III is further supported by the feature marked by arrows in the ac data $\chi$'($H$) shown in Fig. \ref{FieldM}(b). The higher-temperature $T_{N1}$ is smoothly suppressed with increasing field and a related anomaly is not observed at $\mu_{0}H$ = 6 T; this boundary separates phase I and the paramagnetic phase PM.  We note that phase IV may contain an additional phase boundary that is implied by the change in the sign of d$M$/d$T$ upon cooling through the lower-temperature transition in approximately $\mu_{0}H$ = 2.5 T; phase IV may contain both compensated and non-compensated magnetic structures depending on the precise $H,T$.  These results illustrate the need for detailed phase mapping through additional neutron scattering measurements.

Incommensurate magnetic phases can been tuned to a conical phase, a fan phase, or exotic magnetic textures by an external perturbation such as applied pressure, magnetic field or temperature\cite{Muhlbauer2009,Pupal2020,Fab2016, Morin1985}. Indeed, polar tetragonal antiferromagnetic materials have been recently reported to have field-induced complex magnetic textures including metamagnetic textures, skyrmions and merons\cite{Sokolov2019,Kurumaji2017,Pupal2020}. NdCoGe$_3$ with $C_{4v}$ symmetry and a modulated magnetic structure may therefore have the potential to exhibit a topological phase. These results thus promote detailed measurements to reveal similar modulated magnetic phases in related materials, as well as further in-field measurements on NdCoGe$_3$. The outcome from such experiments could provide insight for the effort to understand the relation between topologically nontrivial/trivial magnetic phases in materials with complex magnetic ground states.

\section{Summary}

In summary, NdCoGe$_3$ is a polar tetragonal antiferromagnetic material that crystallizes in a noncentrosymmetric structure type with $C_{4v}$ symmetry. Zero-field specific heat and ac susceptibility measurements as well as neutron diffraction were utilized to reveal a magnetic phase transition at $T_{N2}$ = 3.5 K, below which a complex incommensurate magnetic phase exists (phase II).  The zero-field magnetic phases I and II can be described by the propagation vectors $\vec{k_1}$ = (0.486, 0, 0.385) and $\vec{k_2}$ = (0.494, 0.0044, 0.385), respectively. A unique determination of the magnetic structure of phase II is not possible, though it is observed that even the multi-$\vec{k}$ solutions have non-constant magnetic moments in the modulated structures. The zero field and field-induced phases are summarized in magnetic phase diagrams constructed using physical property data.   For $H \perp c$, field-induced magnetic phases III and IV have been identified, and  the boundary around phase IV appears to be first-order in nature on the basis of hysteresis in the related magnetization data.  These findings reveal that NdCoGe$_3$ deserves further scrutiny as a candidate to host field-induced topological spin textures, and structurally related materials may also be of interest in this regard.

\section{Acknowledgments}

We thank J. M. Perez-Mato, Z. Wang and W. Meier for useful discussions.  This research was supported by the U.S. Department of Energy, Office of Science, Basic Energy Sciences, Materials Sciences and Engineering Division. G.P. and H.S.A were supported by the Gordon and Betty Moore Foundation’s EPiQS Initiative through Grant GBMF4416. This research used resources at the Spallation Neutron Source and High Flux Isotope Reactor, DOE Office of Science User Facilities operated by the Oak Ridge National Laboratory.

\end{document}